\RequirePackage{fix-cm}
\documentclass[natbib,smallextended]{svjour3}   
\smartqed  
\usepackage{graphicx}
\usepackage{mathptmx}     
\usepackage[breaklinks,colorlinks,urlcolor=blue,citecolor=blue,linkcolor=blue]{hyperref}
\newcommand{\gtrsim}%
{\mathrel{\hbox{\rlap{\hbox{\lower5pt\hbox{$\sim$}}}\hbox{$>$}}}}
\newcommand{\lesssim}%
{\mathrel{\hbox{\rlap{\hbox{\lower5pt\hbox{$\sim$}}}\hbox{$<$}}}}
\journalname{Space Science Reviews}
\begin{document}

\title{Origins of the Ambient Solar Wind: Implications for Space Weather}
\titlerunning{Origins of the Ambient Solar Wind}

\author{Steven R.\  Cranmer
\and
Sarah E.\  Gibson
\and
Pete Riley}

\authorrunning{Cranmer, Gibson, and Riley}

\institute{S. R. Cranmer \at
Department of Astrophysical and Planetary Sciences,
Laboratory for Atmospheric and Space Physics,
University of Colorado, Boulder, CO 80309, USA \\
\email{steven.cranmer@colorado.edu}
\and
S. E. Gibson \at
High Altitude Observatory,
National Center for Atmospheric Research,
3080 Center Green Drive, Boulder, CO, 80027, USA \\
\and
P. Riley \at
Predictive Science Inc.,
9990 Mesa Rim Road, Suite 170,
San Diego, CA 92121, USA
}

\date{Submitted: June 2, 2017 / Accepted: August 23, 2017}

\maketitle

\begin{abstract}
The Sun's outer atmosphere is heated to temperatures of millions of
degrees, and solar plasma flows out into interplanetary space at
supersonic speeds.
This paper reviews our current understanding of these interrelated
problems: coronal heating and the acceleration of the ambient solar wind.
We also discuss where the community stands in its ability to forecast
how variations in the solar wind (i.e., fast and slow wind streams)
impact the Earth.
Although the last few decades have seen significant progress in
observations and modeling, we still do not have a complete understanding
of the relevant physical processes, nor do we have a quantitatively
precise census of which coronal structures contribute to specific
types of solar wind.
Fast streams are known to be connected to the central regions of large
coronal holes.
Slow streams, however, appear to come from a wide range of sources,
including streamers, pseudostreamers, coronal loops, active regions,
and coronal hole boundaries.
Complicating our understanding even more is the fact that processes
such as turbulence, stream-stream interactions, and Coulomb collisions
can make it difficult to unambiguously map a parcel measured at 1~AU 
back down to its coronal source.
We also review recent progress---in theoretical modeling, observational
data analysis, and forecasting techniques that sit at the interface
between data and theory---that gives us hope that the above
problems are indeed solvable.
\end{abstract}

\section{Introduction}
\label{sec:intro}

This paper surveys the current state of understanding about how the
solar wind is accelerated along magnetic field lines rooted in the
Sun's hot corona.
It is based on talks and discussions that took place at a June 2016
workshop devoted to {\em The Scientific Foundations of Space Weather}
at the International Space Science Institute (ISSI) in Bern,
Switzerland.
A primary goal of this interdisciplinary workshop was to review
the causal chain of events that link the Sun and the terrestrial
environment, and thus to assess where we stand in our basic physical
understanding of this complex system.
This paper focuses on the origins of the ``ambient'' solar wind,
by which we mean to exclude eruptive events like coronal mass
ejections (CMEs), but to include a wide range of stochastic
processes that produce global-scale structure in the
heliosphere.
This global-scale structure (consisting mainly of fast and slow streams
that interact with one another as they expand out from the Sun)
evolves on timescales from minutes to years, so it is clear that
the term ``ambient'' is not equivalent to ``time-steady.''

The ambient solar wind is known to be a driver of geoeffective
space weather activity.
There are three main ways in which this driving occurs:

\begin{enumerate}

\item

CMEs, the most dramatic source of space weather, accelerate through
a background flow consisting of fast and slow wind streams.
CME flux ropes can be accelerated or decelerated by drag-like
interactions with the surrounding solar wind \citep{Go00,Vr10,Tm11}.
Large-scale spatial structures in the wind can also distort CMEs,
deflect their trajectories, and affect their overall strengths
\citep{Ri97,OP99,Wy04,Is14,ZF17}.
Thus, being able to predict the properties of the ambient solar wind
appears to be a necessary component of predicting CME geoeffectiveness.
In this volume, related reviews of the space weather impacts of CMEs
and other transient forcing events include \citet{issi_manch},
\citet{issi_green}, \citet{issi_eastw}, \citet{issi_mcpher},
\citet{issi_lester}, and \citet{issi_sojka}.

\item

Sustained high-speed wind streams that intersect the Earth's
magnetosphere have been shown to drive geomagnetic activity
\citep[see, e.g.,][]{Ts06,issi_baker,issi_ganus}.
The primary impact of a fast stream appears to be the acceleration
of additional energetic electrons in the radiation belts
\citep[e.g.,][]{Il02,Re03,Jy15,Ki15}.
High-speed streams usually also contain stronger Alfv\'{e}n
waves than the slow wind, and these have been shown \citep{Mg14} to
enhance magnetospheric ultralow frequency (ULF) fluctuations
associated with storms and radiation belt dynamics.

\item

The apparently bimodal structure of the solar wind---i.e., its tendency
to produce fast and slow streams---leads to the production of
compressions, rarefactions, and shocks when the streams interact
with one another.
The passage of such corotating interaction regions (CIRs) past the
Earth's magnetosphere is known to contribute to geomagnetic storm
activity \cite[in this volume, see][]{issi_kilpua,issi_klein}.
Although only a small fraction of the most intense storms appear to
come from CIRs alone \citep{Go91,Hu02,Zh07},
they are responsible for the majority of moderate-strength storms,
especially at solar minimum \citep{Vr11,Ec13}.
CIR events, in combination with high-speed wind streams,
also provide extra heat to the Earth's ionosphere/thermosphere
layers \citep{So09}, which can enhance spacecraft drag and alter
its infrared energy budget.

\end{enumerate}
\noindent
Despite the apparently modest space-weather impacts from fast streams
and CIRs (compared to CMEs) they have the potential for increased
significance because they can persist over long times and are likely to
repeat over multiple solar rotations \citep[see, e.g.,][]{SR97,BD06}.

\begin{figure*}
\includegraphics[width=1.01\textwidth]{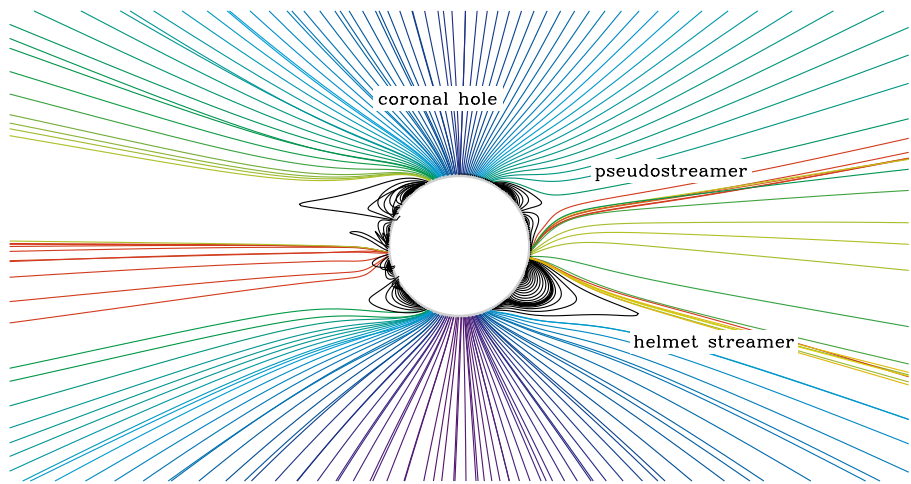}
\caption{Closed (black) and open (multi-color) magnetic field lines
traced from a time-steady solution of the polytropic MHD conservation
equations, computed by the Magnetohydrodynamics Around a Sphere (MAS)
code \citep{Li99}.
Photospheric boundary conditions were from Carrington Rotation
2058 (June--July 2007).
Colors of open field lines correspond to the \citet{WS90} expansion
factor: $f \leq 4$ (violet), $f \sim 6$ (blue), $f \sim 10$ (green),
$f \sim 15$ (gold), $f \geq 40$ (red).
Labeled structures are discussed in more detail in
Sect.\  \ref{sec:obs:corona}.}
\label{fig01}
\end{figure*}

An ongoing topic of debate is whether the solar wind is truly
bimodal (i.e., cleanly separable into two distinct source regions).
In the half-dozen years around each minimum in the Sun's 11-year
activity cycle, there are large unipolar coronal holes at the north and
south poles, with mostly closed fields at low latitudes.
Figure \ref{fig01} shows extrapolated field lines from a
rotation-averaged magnetohydrodynamic (MHD) model constructed for
a representative solar-minimum time period.
We have high confidence that the fast solar wind is rooted in the
central regions of coronal holes.
The slow solar wind appears to be associated with ``everywhere else''
on the Sun that connects out to the distant heliosphere.
Some slow-wind source regions may start as closed magnetic loops
and undergo jet-like magnetic reconnection.
Other regions may be topologically similar to fast-wind source regions,
but with lower levels of momentum and energy deposition.
This paper will discuss several unanswered questions about the solar
wind's bimodality, magnetic topology, and radial evolution.

The remainder of this paper is organized as follows.
Sect.\  \ref{sec:obs} provides a review of solar wind observations,
both remote and {\em in~situ,} as well as a discussion of how coronal
structures appear to be connected to their counterparts in the
heliosphere.
In Sect.\  \ref{sec:model} we summarize the current state of
theoretical solar wind modeling.
Sect.\  \ref{sec:paths} gazes into the crystal ball to speculate
about what future improvements are needed, and
Sect.\  \ref{sec:conc} concludes with some broader context about
the impact of this work on other fields.
Because the solar wind has been studied by hundreds of researchers
for more than a half-century, this paper cannot be truly comprehensive
in its review of the literature.
Interested readers are urged to fill in the gaps by surveying other
reviews, such as those by \citet{Ds67}, \citet{HA70}, \citet{Hu72},
\citet{Le82}, \citet{Ba92}, \citet{P97}, \citet{Cr02,Cr09},
\citet{Ma06}, \citet{Ve10}, \citet{Ab16}, and \citet{Ch16}.

\section{Observations of Solar Wind Origins}
\label{sec:obs}

In order to identify the physical processes responsible for producing
the solar wind, we must have accurate empirical measurements of the
plasma and field properties.
Sect.\  \ref{sec:obs:1AU} summarizes {\em in~situ} interplanetary
measurements, and Sect.\  \ref{sec:obs:corona} describes
remote-sensing observations of the coronal origin regions near the Sun.
Sect.\  \ref{sec:obs:mapping} discusses how periodicities and other
correlations between data sets have been used to improve our
understanding of ``what connects to what'' between the corona and
heliosphere.

\subsection{Interplanetary Measurements}
\label{sec:obs:1AU}

Evidence for the existence of an outflow of ``corpuscular radiation''
(i.e., charged particles) from the Sun accumulated gradually throughout
the early 20th century \citep[see historical reviews by][]{Ds67,Hu72}.
Early {\em in situ} detections of solar wind particles were made
between 1959 and 1961 by Russian and American spacecraft that left
Earth's magnetosphere.
The continuous, supersonic, and possibly bimodal nature of the solar
wind was confirmed by {\em Mariner 2} on its journey to Venus
\citep{NS62}.
Those early data indicated a range of outflow speeds (roughly from
250 to 800 km~s$^{-1}$) that seem to act as an organizing quantity.
In other words, many of the other plasma and field quantities measured
at 1~AU appear to be correlated with whether one is in a fast or slow
stream.

Table \ref{table01} summarizes some representative properties of
the fast and slow wind regimes as revealed over the past half-century
of exploration \citep[see also][]{Sh06}.
There is still substantial debate about whether the solar wind plasma
can be classified into more than two distinct types---based on, e.g.,
source regions, acceleration mechanisms, or local plasma physics---and
whether or not the wind speed is in fact a reliable indicator of
which type is being detected \citep[see, e.g.,][]{Wa09,Zu12,Sk15,Ne16}.
Difficulties arise because much of the solar wind at 1~AU has undergone
some kind of processing or mixing (see Sect.\  \ref{sec:obs:mapping}),
such that the global magnetic topology and coronal connections are
not easy to determine.

\begin{table}
\renewcommand\thetable{1}
\caption{Properties of slow and fast solar wind streams.}
\label{table01}
\begin{tabular}{lcc}
\hline\noalign{\smallskip}
\multicolumn{1}{c}{Quantity} &
\multicolumn{1}{c}{Slow wind} &
\multicolumn{1}{c}{Fast wind} \\
\noalign{\smallskip}\hline\noalign{\smallskip}
Radial flow speed &
250--450 km s$^{-1}$ & 450--800 km s$^{-1}$ \\
Proton density (1 AU) &
5--20 cm$^{-3}$ & 2--4 cm$^{-3}$ \\
Proton temperature (1 AU) &
0.03--0.1 MK & 0.1--0.3 MK \\
Electron temperature (1 AU) &
0.1--0.15 MK & $\sim$ 0.1 MK \\
Freezing-in temperature (corona) &
1.4--1.7 MK & 1.0--1.3 MK \\
Helium abundance &
0.5\%--4\% & 3\%--5\% \\
Heavy ion abundances &
low-FIP enhanced & $\sim$ photospheric \\
Ion/proton temperature ratio &
$< \, m_{\rm ion}/m_p$ & $> \, m_{\rm ion}/m_p$ \\
Coulomb collisional age (1 AU) &
0.1--10 & 0.001--0.1 \\
Coronal WSA expansion factor &
15--100 & 3--10 \\
Coronal sources (Sect.\  \ref{sec:obs:corona}--\ref{sec:obs:mapping}) &
streamers, quiet loops, active regions, &
 coronal hole cores \\
 & coronal hole boundaries, separatrices & \\
\noalign{\smallskip}\hline
\end{tabular}
\end{table}

Despite the above difficulties, there are many regularities in the
{\em in~situ} data.
The raw probability distribution of wind speeds $u$ in the ecliptic
is usually single-peaked around 400 km~s$^{-1}$, with a relatively
sharp cutoff below about 250 km~s$^{-1}$ and a skewed tail toward
higher speeds \citep[e.g.,][]{Go71,Mg11a}.
An interesting exception was in 2008 during the ``peculiar solar minimum''
when the presence of long-lived, low-latitude coronal holes led to a
truly bimodal distribution of solar wind speed at 1~AU \citep{dT11}.
Proton and electron densities $n$ are negatively correlated with speed,
but the mass flux (i.e., the product $nu$) has a slight residual trend
toward higher values in the slow wind.
\citet{Le12} found that the kinetic energy flux (proportional to $nu^3$)
is very nearly constant as a function of wind speed, latitude, and
solar cycle.
The radial magnetic flux also tends to be reasonably constant throughout
the low- and high-latitude heliosphere \citep{SB95}, but its overall 
value does change as a function of global solar activity \citep{SC07}.
Any theoretical model of the solar wind must reproduce these trends
and quasi-invariants.

Heliospheric measurements in the ecliptic plane tend to show a
preponderance of slow solar wind, with high-speed streams being
occasional interlopers.
This led to early widespread identification of the slow wind as the
``ambient'' background state \citep[e.g.,][]{Hu72}.
However, there were hints---going back to at least \citet{Ba77}---that
the fast wind was a much better candidate for being the most time-steady
and quiescent type of solar wind.
The {\em Ulysses} probe confirmed this picture when it left the
ecliptic plane and showed that the fast wind is ubiquitous over large
polar coronal holes, which (1) persist over more than half of each solar
cycle, and (2) expand out to fill the majority of the heliospheric volume
\citep{Go96,Ms01,Mc08}.

In the 1990s, {\em Ulysses} and {\em ACE} also began to show that ion
composition measurements (i.e., both elemental abundances and ionization
states) can be used to reliably distinguish slow and fast wind streams
from one another.
These composition signatures are established close to the Sun and are
subsequently ``frozen in'' along most of the extent of each wind stream.
On the other hand, the wind speed itself continues to evolve dynamically
between the Sun and 1~AU as streams interact with one another.
Thus, ion composition is suspected to be more reliable as a
wind-stream identification tag than the flow speed
\citep[see, e.g.,][]{Ne16,Fu17}.
The ratio of O$^{+7}$ to O$^{+6}$ charge-state number densities tends to
be the most widely reported composition signature, mainly because the
large oxygen abundance allows for good measurement statistics.
However, \citet{La12a} suggested that the relative fractions of carbon
ions C$^{+4}$, C$^{+5}$, and C$^{+6}$ may be more precise probes of the
plasma conditions in the low corona ($r \approx 1.2 \, R_{\odot}$)
where the freezing-in occurs.

\begin{figure*}
\includegraphics[width=1.01\textwidth]{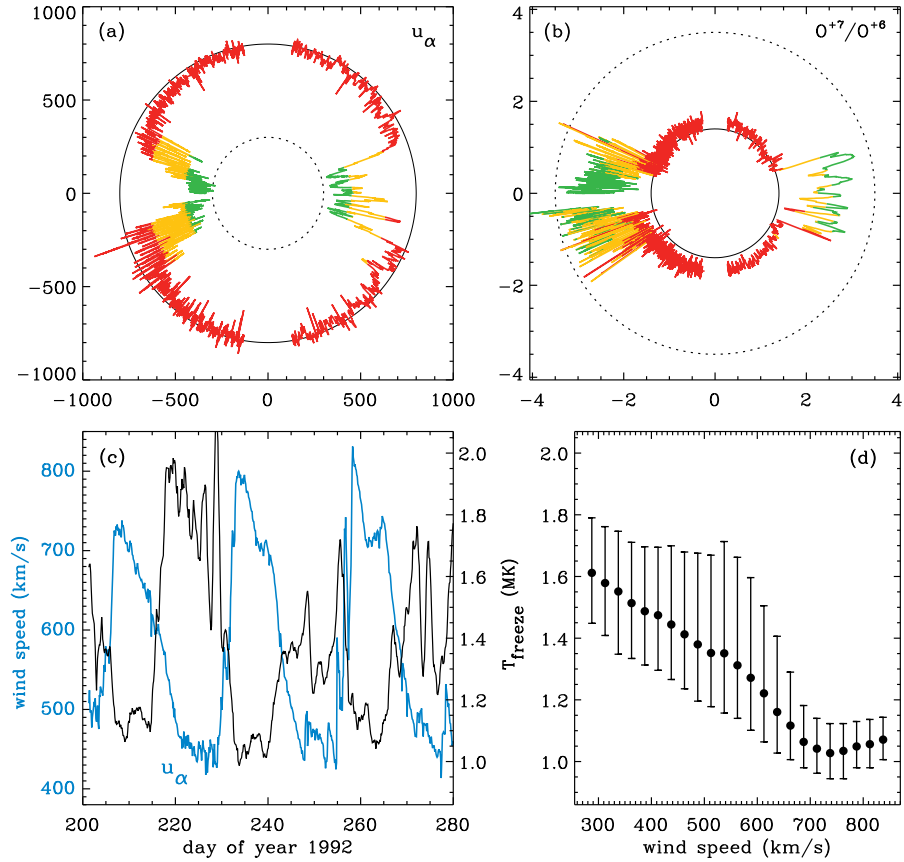}
\caption{Polar plots of (a) alpha particle wind speeds, and (b) ratios
of O$^{+7}$ to O$^{+6}$ ion number densities from {\em Ulysses}/SWICS
during its first high-latitude orbit in 1992--1997.
Parcels are color-coded by wind speed (green: $u < 450$ km~s$^{-1}$,
red: $u > 650$ km~s$^{-1}$, yellow: intermediate) with the same
labels applied to data points in panel (b).
The speed/ion-ratio anticorrelation is also shown in (c) as a
function of time for several solar rotations in 1992
\citep[i.e., the same period analyzed by][]{Ge95}.
Panel (d) shows the same anticorrelation, collected into 25 km~s$^{-1}$
bins over the entire {\em Ulysses} mission, with each bin's median
(filled circles) and $\pm 1\sigma$ error bars.
In (c) and (d) the ion ratio was converted to freezing-in temperature
(see text).}
\label{fig02}
\end{figure*}

Figure \ref{fig02} shows data from the {\em Ulysses} SWICS (Solar Wind
Ion Composition Spectrometer) instrument final archive to illustrate how
the traditional O$^{+7}$/O$^{+6}$ charge-state ratio varies as a
function of solar wind speed \citep{Gl92,vS00}.
The polar plots show (a) wind speed in km~s$^{-1}$, and (b) a scaled
ratio with magnitude $3.4 + \log_{10}(\mbox{O}^{+7} / \mbox{O}^{+6})$,
as a function of latitude during an orbit near solar minimum.
In panels (c)--(d), the equivalent O$^{+7}$/O$^{+6}$ freezing-in
temperature (i.e., the electron temperature corresponding to a given
charge-state ratio in coronal equilibrium) was computed from ionization
balance curves provided in version 7.1 of CHIANTI \citep{La12b}.
Note that panel (c) sometimes indicates abrupt changes in the ionization
state at intermediate wind speeds, but panel (d) shows that,
statistically speaking, the trend is rather gradual.

Additional clues about the physical origins of fast and slow wind
streams come from the kinetic properties of the plasma.
It has been known since the first decade of interplanetary exploration
\citep[e.g.,][]{SH66} that solar wind parcels are not just expanding
adiabatically, but are continuing to undergo changes in their energy
budgets at 1~AU and beyond.
The relatively slow radial decline in particle temperature $T(r)$
indicates some combination of sustained thermal energy input
(a continuation of coronal heating) and strong heat conduction due to
the presence of skewness in the velocity distributions.
The latter is certainly true for electrons \citep[e.g.,][]{Ba13},
and it has been recently argued to be an important contributor to proton
thermodynamics as well \citep{Sc15}.

\begin{figure*}
\hspace*{0.05in}
\includegraphics[width=0.97\textwidth]{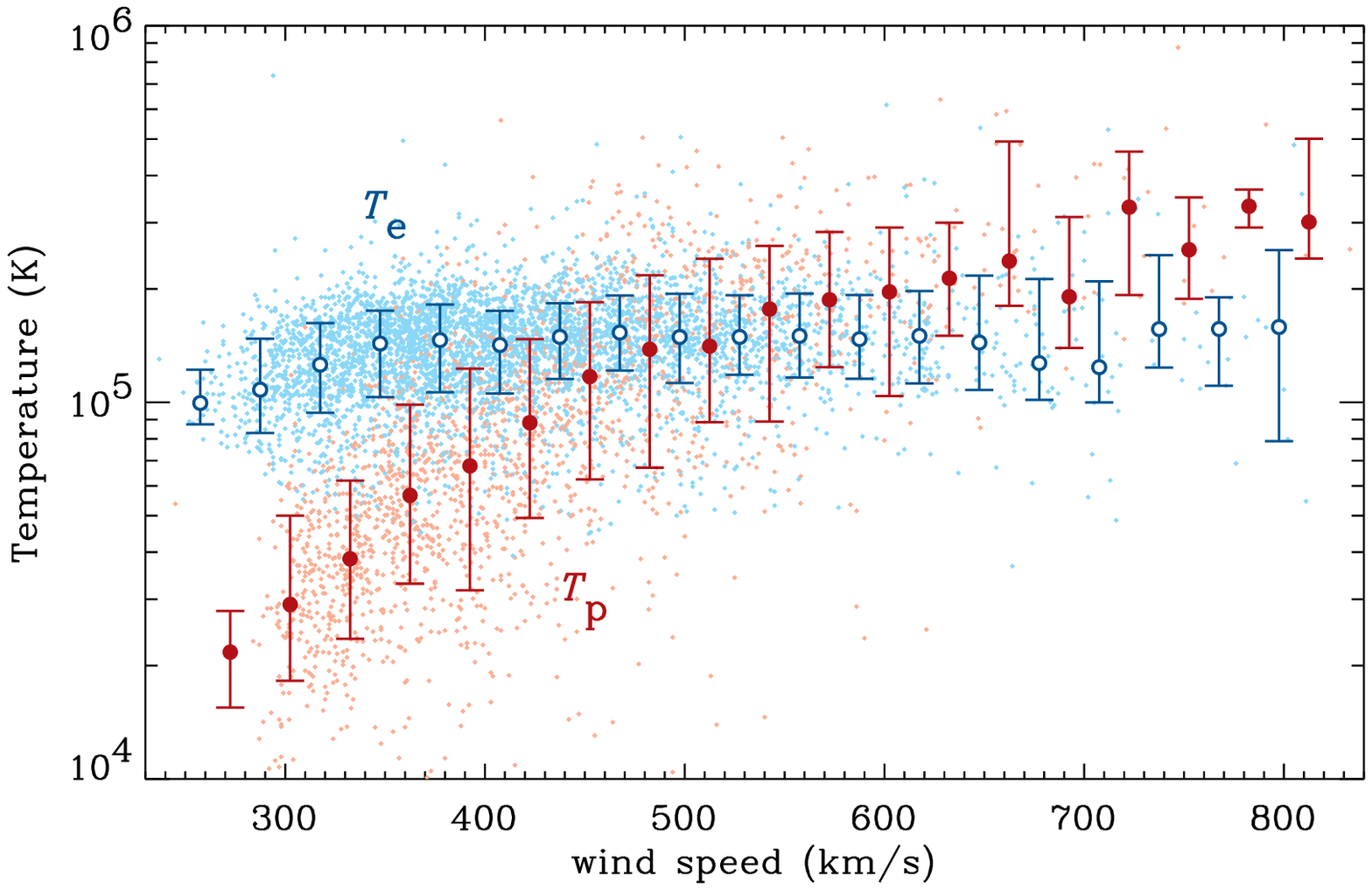}
\caption{Hourly averaged proton (red) and electron (blue) temperatures
measured at 1~AU by {\em ISEE--3} \citep{Nw98} between
January 1980 and October 1982.
Small points indicate individual measurements, and large symbols with
error bars show median and $\pm 1 \sigma$ values within 30 km~s$^{-1}$
bins of solar wind speed.}
\label{fig03}
\end{figure*}

Coulomb collisions in the solar wind appear to be infrequent enough
to allow the protons and electrons to evolve away from a common
thermal state.
Figure \ref{fig03} illustrates this by showing the dominant trends of
proton temperature $T_p$ and electron temperature $T_e$ versus wind
speed at 1~AU.
The protons appear to be strongly correlated with wind speed
\citep[see also][]{El12} while the electrons are much less sensitive
to local conditions.
In the slow wind, it seems possible that stronger electron
conduction keeps the coronal $T_e$ high for a larger range of distance,
while weaker proton conduction (and a lack of equilibrating collisions)
allows the protons to cool off more rapidly \citep[see, e.g.,][]{Fr88}.
In the fast wind, the data show $T_p > T_e$, which suggests sustained
heating for the protons.
There have been several empirical estimates of heat input rates
that indicate the protons receive more ``extended  coronal heating''
than do the electrons \citep{St09,Cm09,Sp15}.

Protons in the inner heliosphere also tend to exhibit thermal
anisotropies, with unequal temperatures measured perpendicular and
parallel to the background magnetic field \citep{Ma06}.
In the fast solar wind, the proton magnetic moment
$\mu \propto T_{\perp}/B$ has been seen to increase with increasing
heliocentric distance \citep{Ma83}.
This suggests the existence of kinetic wave-particle interactions
that transfer thermal energy to only {\em some} of the
proton degrees of freedom.
When the proton data at 1~AU are plotted in a two-dimensional plane
of the anisotropy ratio (${\cal R} = T_{\perp}/T_{\parallel}$) versus
the parallel plasma beta parameter ($\beta_{\parallel}$, parallel gas
pressure divided by magnetic pressure), the resulting distribution
of data points \citep[see][]{He06,Mu12}
provides additional constraints on the nature of
wave-particle interactions that energize the protons.
Some kinds of simple linear theory---i.e., the damping of a cascading
spectrum of ion cyclotron waves \citep{Cr14b}---predict reasonably
correct shapes for the populated region in
(${\cal R},\beta_{\parallel}$) parameter space.
However, more physically realistic numerical simulations
\citep[e.g.,][]{Sv15,He17} may be needed to reproduce all of the
relevant details of this region.

Ions heavier than hydrogen are also useful probes of kinetic physics
in the collisionless solar wind.
Both alpha particles and other minor ion species are heated and
accelerated preferentially in comparison to the protons.
At 1~AU, these differences appear to be organized by the Coulomb
collisional ``age'' of the solar wind parcel; i.e., parcels that
experience the fewest number of collisions between the Sun and 1~AU
show the strongest departures from thermal equilibrium
\citep[e.g.,][]{Ka08}.
Preferential ion heating appears to be necessary condition for
preferential ion acceleration.
\citet{Ge70} investigated models without extra heating and found that
ions tend to flow out more slowly than the protons; in fact, in those
models Coulomb friction may help bring ions up to the proton outflow
speed, but no faster.
\citet{RA75} and others realized that heating the ions more strongly
than the protons---at least proportionally to their masses (to provide
comparable pressure gradients) or even more than that (to accelerate
them even faster)---was a natural explanation for the data.

\begin{figure*}
\includegraphics[width=1.01\textwidth]{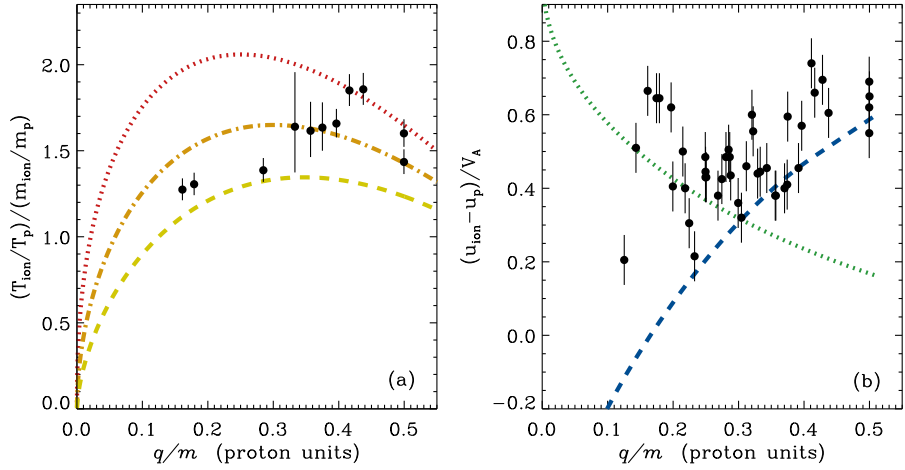}
\caption{Heavy-ion preferential heating (a) and acceleration (b) with
respect to solar wind protons at 1~AU.
Both panels show relative ion--proton quantities versus the charge/mass
ratio ($q/m$) in units of proton charge/mass.
(a) Points show the ratio of ion to proton squared thermal speeds
($v_{\rm th}^2 \propto T/m$) from {\em{ACE}}/SWICS \citep{Tra16}.
(b) Points show the difference between ion and proton bulk flow speeds
in units of the local Alfv\'{e}n speed, as measured by 
SWICS and SWEPAM on {\em ACE} \citep{Bg11}.
For discussion of the model curves, see text.}
\label{fig04}
\end{figure*}

Figure \ref{fig04} shows recent measurements of preferential ion
heating \citep{Tra16} and preferential ion acceleration \citep{Bg11}
measured at 1~AU for collisionally young plasma that tends to be
dominant in the fast wind.
The particles measured by {\em ACE} include multiple ionization stages
of He, C, N, O, Ne, Mg, Si, S, Ca, and Fe.
Each ion temperature is shown as a squared thermal speed
(i.e., $T_i / m_i$) in units of a similar quantity corresponding to
the protons.
The ion bulk flow speeds are shown as differences ($u_i - u_p > 0$) in
units of the local Alfv\'{e}n speed $V_{\rm A}$.

There is still no consensus about the identity of the physical
processes responsible for the observed ion properties.
The model curves shown in Figure \ref{fig04} are meant to illustrate
the challenges inherent in explaining the data with a single
kinetic theory.
Curves in Figure \ref{fig04}a are predictions from ion
cyclotron resonance excited by MHD turbulence
\citep[see equation 26 of][]{Cr02}.
These curves correspond to a turbulent power-law spectrum
$P \propto k_{\parallel}^{-\eta}$ with $\eta = 1.57$ (red dotted curve),
$\eta = 1.47$ (orange dot-dashed curve), and
$\eta = 1.37$ (yellow dashed curve), where $k_{\parallel}$ is the
wavenumber of cyclotron resonant fluctuations in the direction parallel
to the background magnetic field.
The model curves in Figure \ref{fig04}b are upper and lower limits on
the differential ion flow speeds compatible with ion cyclotron resonance.
The cyclotron waves were assumed to obey a cold-plasma dispersion
relation \citep[e.g.,][]{HI02} with alpha particles flowing
$0.55 V_{\rm A}$ faster than protons, as measured by \citet{Bg11}.
The blue dashed curve shows minimum resonant ion speeds for
$k_{\parallel} < 0$, and the green dotted curve shows maximum resonant
ion speeds for $k_{\parallel} > 0$ \citep[see also][]{MM82}.
It is important to note that the relevance of these ion-cyclotron
curves to the data has not yet been demonstrated conclusively.
However, it may be noteworthy that the
rightmost ``wedge'' region of the plot (below the blue curve and
above the green curve, for $q/m > 0.3$) is firmly excluded by both
curves and is also more or less empty of data points.

\subsection{Coronal Measurements}
\label{sec:obs:corona}

A wide variety of remote observation techniques---direct imaging,
spectroscopy, radio sounding, and coronagraphic occultation---have
been used to put useful constraints on solar wind origins
\citep{BE90,Ko06,Hb13,Ju13,Sz14}.
These techniques have been implemented on a number of different
platforms---spacecraft, rockets, ground-based observatories, and
movable ``eclipse-chasing'' instruments---each with its own
unique advantages and challenges.
The combined analysis of data from these different platforms (also
including {\em in~situ} particle and field detection) has been a
crucial ingredient in the advances made so far in our knowledge about
the complex Sun-heliosphere system.

The solar disk contains small-scale features (e.g., bright points,
faculae, ephemeral regions) and medium-scale structures (e.g., active
regions, filaments) that are associated mainly with closed
magnetic loops.
Because these features do not appear to be connected continuously to
the open heliosphere, much of the work in studying solar wind origins
has focused on {\em large-scale} features such as coronal holes
and streamers.
The remainder of this subsection describes these features.
However, Sect.\  \ref{sec:model:debate} discusses a range of proposed
coronal heating processes that includes the dynamical evolution of
(temporarily) closed magnetic regions.

\begin{figure*}
\includegraphics[width=1.01\textwidth]{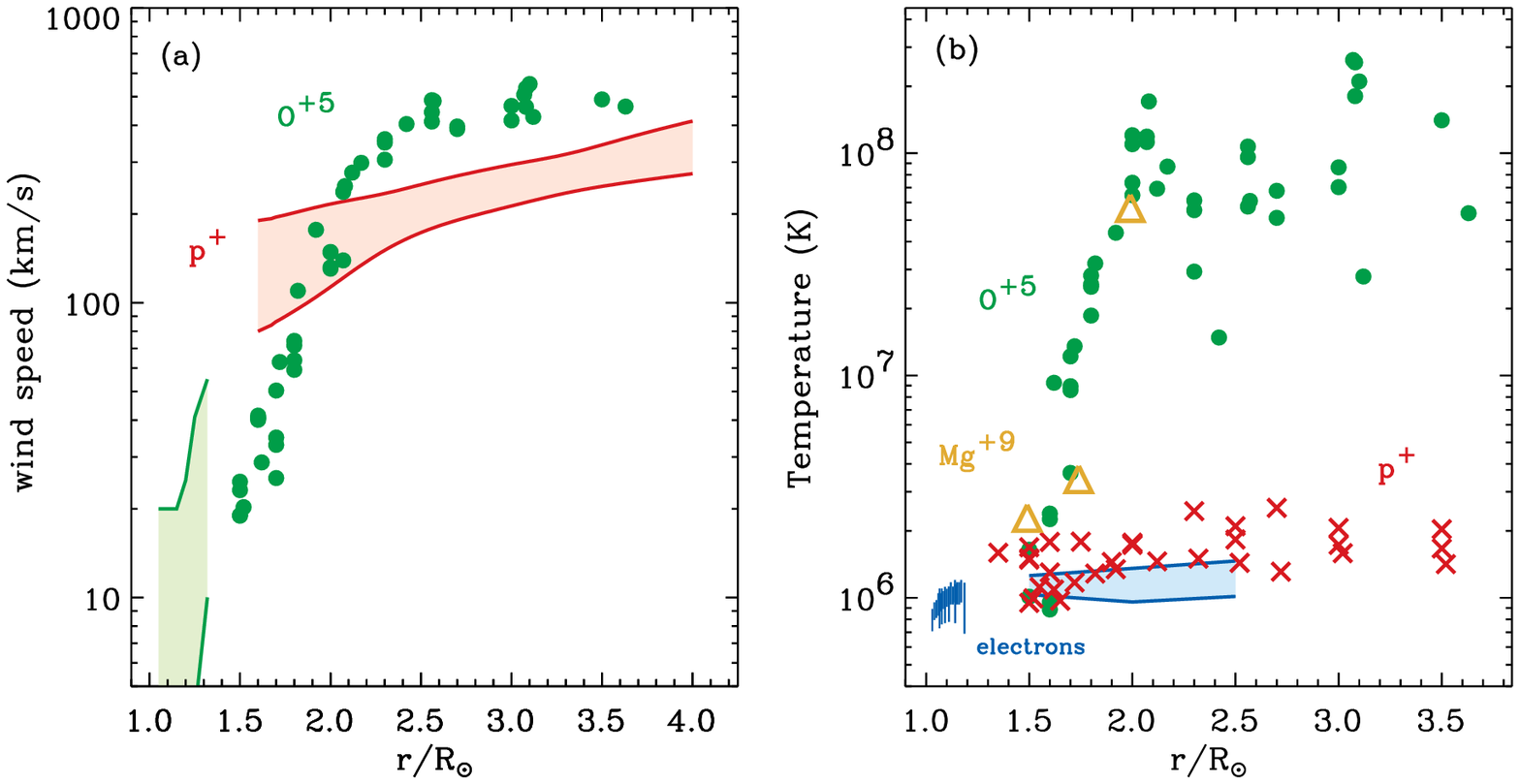}
\caption{Off-limb measurements of (a) outflow speeds and
(b) temperatures above polar coronal holes.
Red: proton flow speeds \citep{Cemp} and temperatures assembled from
various sources \citep[e.g.,][]{An00,Cr09}.
Green: O$^{+5}$ data points from \citet{Cr08}, and near-Sun
bounded region in outflow speed from \citet{Te03}.
Gold: Mg$^{+9}$ ion temperatures \citep{Ko99}.
Blue: electron temperatures at $r < 1.3 \, R_{\odot}$ \citep{La08}
and $r > 1.5 \, R_{\odot}$ (Cranmer 2017, in prep).
See text for details, and original sources for error bars.}
\label{fig05}
\end{figure*}

Coronal holes are low-density patches of nearly unipolar magnetic
flux on the surface that appear to expand out superradially into
the heliosphere.
The central regions of large coronal holes are known sources of fast
solar wind \citep{Wi68,KTR73,No73}.
Because they are associated with tenuous, collisionless plasmas
and are long-lived time-steady structures, coronal holes have been
ideal hunting grounds for similar kinetic effects as seen in fast
wind streams at 1~AU.
Figure \ref{fig05} summarizes the evidence found by the
Ultraviolet Coronagraph Spectrometer (UVCS) instrument on the
{\em Solar and Heliospheric Observatory} ({\em{SOHO}}) for
preferential ion heating and acceleration above coronal holes
\citep[see, e.g.,][]{Ko06}.
Most of this evidence comes from the comparison of proton properties
(measured by proxy using the neutral hydrogen H~I Ly$\alpha$ line)
and O$^{+5}$ ions (similarly probed by the O~VI 103.2--103.7 nm
resonance doublet).

The flow speeds in Figure \ref{fig05}a were derived from the
so-called ``Doppler dimming'' technique, which takes advantage of
the fact that fewer solar-disk photons are scattered into our
line of sight when the atoms (i.e., the coronal scattering centers)
are Doppler shifted away from the narrow spectral window of the
available photons.
The ion temperatures in Figure \ref{fig05}b were derived from
spectral line widths and associated modeling, and are mainly probes
of $T_{\perp}$.
These temperatures were corrected to remove nonthermal line
widths associated with MHD waves and turbulence.
This is a model-dependent correction, but it is based on additional
observational data (see below).
The electron temperatures at $r > 1.5 \, R_{\odot}$ are preliminary
results from an empirical generalization of older hydrostatic
scale-height techniques \citep[see, e.g.,][]{LS16} using the UVCS
visible-light and Ly$\alpha$ data as constraints.
These estimates of $T_e$ generally agree with existing visible-light
Thomson scattering results \citep{Rg11}.

The remote-sensing data provide evidence for $T_{\rm ion} \gg T_p > T_e$
in the source regions of fast solar wind, which is reminiscent of
the heliospheric data at distances greater than 0.3~AU.
The initial reports of heavy ions with temperatures of order $10^8$~K
(i.e., even hotter than in the solar core), together with
$T_{\perp} > T_{\parallel}$ and ion flow speeds roughly double those
of the protons at $r \approx 3 \, R_{\odot}$, were surprising.
There was some skepticism about the uniqueness of these interpretations
of the data \citep[e.g.,][]{RS04}.
However, more rigorous data analysis \citep{Cr08} has
generally upheld those initial results, albeit with some tempering
(i.e., the O$^{+5}$ anisotropy ratio $T_{\perp}/T_{\parallel}$ was
found to be more like 3--10, instead of the earlier claim of $\sim$100).

In addition to coronal holes, the large-scale corona contains a
variety of other magnetic features that appear to be connected to
the slow solar wind (see Sect.\  \ref{sec:obs:mapping}).
When observing above the solar limb, the most striking of these are
the bright streamers illustrated in Figure \ref{fig01}.
The magnetic field in streamers appears to be closed at low heights,
with surrounding open field lines converging above a cusp-like point
at the top.
The helmet-like appearance of many streamers has been compared to
the 19th century Prussian {\em pickelhaube,} and it is clear that the
solar wind acts to open up the magnetic field above these structures.
Streamers are generally assumed to be sources of low-speed solar wind,
but the precise topological connections (Sect.\  \ref{sec:obs:mapping})
and mass-release mechanisms (Sect.\  \ref{sec:model:debate})
are still being debated.

In recent years, distinctions have been made between:
(1) helmet streamers that expand up from a bipolar loop, and thus have
a large current sheet between the two opposite-polarity legs, and
(2) ``pseudostreamers'' that are connected to an even number of
bipoles, and thus have legs with the same polarity
\citep[e.g.,][]{Wa07,RL12,Rm14}.
There are also differences in the plasma properties between large
quiescent equatorial streamers and the brighter, more compact
streamers associated with active regions \citep{Lw01,Ko02}.
The relatively high densities seen in all coronal streamers
\citep{Gi99,St02} appear to indicate rapid Coulomb collisions that
generally lead to temperature equilibration ($T_p \approx T_e$).
However, the largest streamers do start to exhibit collisionless
kinetic effects, such as high O$^{+5}$ temperatures similar to
what is seen in coronal holes, above their cusps \citep{Fr03}.

\begin{figure*}
\includegraphics[width=1.01\textwidth]{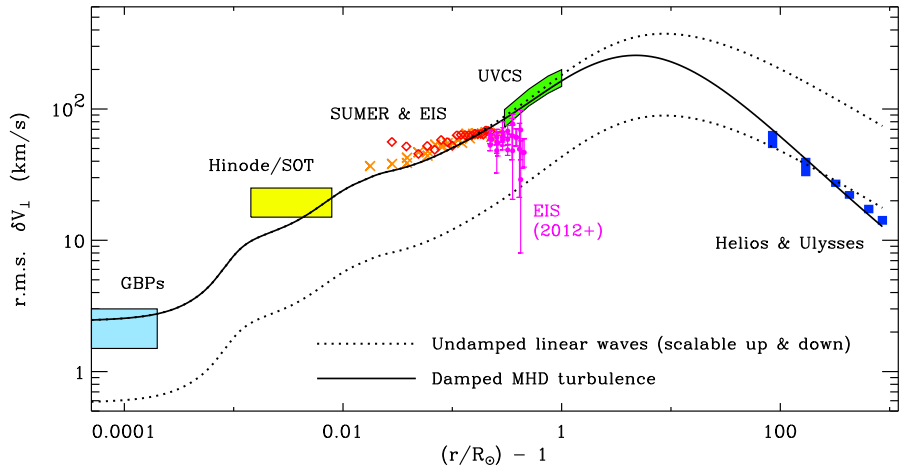}
\caption{Height dependence of transverse velocity amplitudes of
MHD fluctuations in coronal holes and the fast solar wind.
Model curves and the photospheric G-band Bright Point (GBP) data
are from \citet{CvB05}.
Other data, from left to right, are from
Type II spicule motions observed by {\em Hinode}/SOT \citep{DP07},
nonthermal line broadening from SUMER, EIS, and UVCS (see text),
and direct {\em in~situ} measurement from {\em Helios} and
{\em Ulysses} \citep{Bv00}.}
\label{fig06}
\end{figure*}

Both coronal holes and streamers are intrinsically time-variable.
In addition to changes in connectivity that occur as the Sun's
magnetic field evolves over multiple solar rotations, the corona is
also observed to be full of large-amplitude oscillations
(e.g., waves, shocks, and transient eddies).
A comprehensive review of oscillation measurement techniques is
beyond the scope of this paper \citep[see, e.g.,][]{Cr02,Cr04,Na06},
but there are several aspects that are relevant to solar wind origins:

\begin{enumerate}

\item

Sensitive measurements of the off-limb coronal intensity
allow {\em low-frequency density fluctuations} to be tracked in
space and time.
The tips of most helmet streamers appear to be unstable to the
production of blob-like plasmoids that flow out with the slow
solar wind \citep{Sh97,Sh09,Wa00,Pk16}.
Similar features continue to be detected as 1--2 hour density
modulations at larger distances \citep{VV15}, and they appear to
be most intense in the heliospheric current sheet (HCS).
Above coronal holes, there are appear to be weak, but ubiquitous
field-aligned compressive waves with periods of order 10--20 minutes
\citep{Of99,Th13,Liu15}.

\item

Radio telescopes probe plasma properties near the Sun by measuring how
signals are distorted by changes in the refractive index as they pass
through the corona \citep[e.g.,][]{Ba01}.
Interplanetary scintillation (IPS) measurements are sensitive to
{\em high-frequency density fluctuations} (i.e., millisecond timescales),
and additional information can be extracted about the coronal
magnetic field and the solar wind speed.
Global IPS maps of solar wind acceleration show the presence of fast
and slow streams \citep{KK90,Gr96,Im14}, but some information is lost
by the integration over long lines of sight.
\citet{Ef10} detected spatial anisotropy in radio-detected turbulent
eddies at heliocentric distances smaller than $\sim$25~$R_{\odot}$,
and isotropy above $\sim$30~$R_{\odot}$.
This is a similar qualitative transition as the one seen in the shapes
of larger visible-light structures resolved by heliospheric imagers.
However, for the latter, \citet{Df16} found that the transition to
isotropy does not occur until at least 60--80 $R_{\odot}$.

\item

A combination of motion-tracking and spectroscopic Doppler-shift
techniques allows {\em transverse Alfv\'{e}nic fluctuations} to be
detected in the solar wind.
It is suspected (see Sect. \ref{sec:model:debate}) that
Alfv\'{e}n waves and turbulence are major players in heating the
extended corona and solar wind.
Figure \ref{fig06} shows a summary of inferred velocity amplitudes
over polar coronal holes.
The associated model curves show predictions for undamped and damped
Alfv\'{e}nic turbulence from \citet{CvB05}.
Measured amplitudes derived from nonthermal line widths are shown from
SUMER/{\em{SOHO}} \citep[][orange crosses]{Bj98},
near-limb EIS/{\em{Hinode}} data \citep[][red diamonds]{LC09},
and UVCS/{\em{SOHO}} \citep[][green region]{Es99}.
Taken together, those data appeared to agree well with the
predictions for Alfv\'{e}n waves that dissipate and heat the corona.
More recently, however, the EIS instrument has been used to probe
larger heights above the poles; magenta points show data from
\citet{Hh13} \citep[see also][]{Hh12,BA12,Gu17}.
There is now clearly some ``tension'' with the model curves
and with the inferred UVCS result from \citet{Es99}.

\end{enumerate}
\noindent
Figure \ref{fig06} makes it clear that our knowledge of the global
evolution of waves and turbulence in the solar wind is still lacking.
The recent EIS data call into question our understanding of where
Alfv\'{e}n waves are damped and how their energy is converted to heat.
There is also some inherent uncertainty in interpreting the properties
of off-limb emission lines, especially when observing diffuse areas
such as coronal holes.
If the line of sight contains $N$ independently fluctuating flux tubes,
with $N \gg 1$, then many of the desired diagnostics (e.g., Doppler
shifts or plane-of-sky swaying motions) are reduced in amplitude by
roughly $1/\sqrt{N}$.
Monte Carlo forward models \citep{DP07,Mc11} have proven to be helpful
in estimating the magnitude of this effect, but definitive ``inversions''
are not yet possible.
In Sect.\  \ref{sec:paths}, we discuss future efforts to improve upon
the existing measurements.

\subsection{Periodicities Linking the Sun and Heliosphere}
\label{sec:obs:mapping}

There is not yet a fully-understood one-to-one mapping between
observed features in the corona and {\em in~situ} detected structures
in the heliosphere.
Multi-point measurements made over multiple solar rotations---sometimes
extending to multiple solar cycles---have helped us find correlations
between large, long-lived structures on the Sun and in the solar wind.
Whether or not these correlations are related to physics-based causations
is a separate issue, but good correlations provide good starting points
for space weather prediction.

For example, the half-century long OMNI database of plasma and field
measurements at 1~AU has been shown to be useful for long-baseline
studies of all kinds \citep[e.g.,][]{OM00,KP05,Lc09}.
Analogous databases for regions near the Sun have run the gamut from
careful hand-drawings based on daily images \citep{HR02,Mp03} to
automated ``big data'' feature-extraction systems \citep{Mr12,Bb14}.
Taking inspiration from worldwide events like the 1957 International
Geophysical Year, multiple communities came together in coordinated
projects---e.g., three ``Whole Sun Months'' in 1996, 1998, and 1999
\citep{GK99,Ri99,Br00} and a ``Whole Heliosphere Interval'' in 2008
\citep{Gi09,Gi11,Ri11,Th11}---to improve our understanding of
Sun-heliosphere connectivity.

\begin{figure*}
\hspace*{0.05in}
\includegraphics[width=0.95\textwidth]{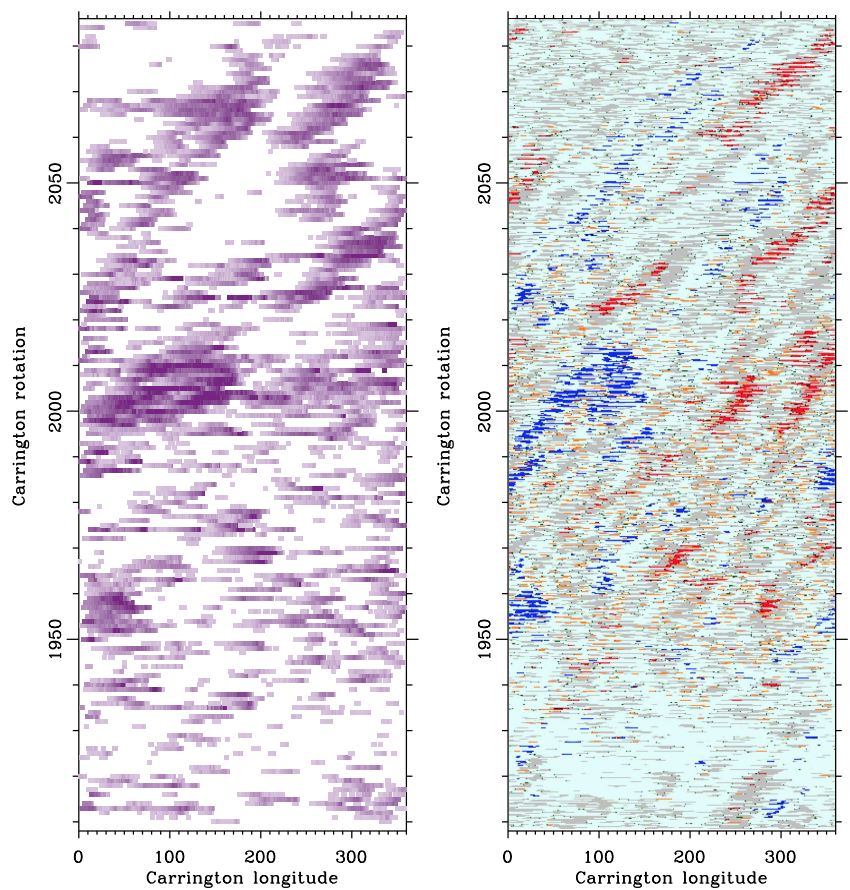}
\caption{Carrington rotation stack plots showing (a) in-ecliptic OMNI
wind speeds and (b) surface features from the McIntosh archive,
both for the duration of solar cycle 23
(June 1996 to July 2009).
In panel (a), white denotes $u \leq 450$ km~s$^{-1}$ and increasingly
darker shades of purple eventually saturate at the darkest color
for $u \geq 750$ km~s$^{-1}$.
Longitudes have been offset by 50.55$^{\circ}$, or 3.83 days, to account
for propagation from the Sun to 1~AU at a mean speed of 450 km~s$^{-1}$.
Panel (b) shows equatorial ($\pm 20^{\circ}$ from equator) features,
with blue [red] showing coronal holes of positive [negative] polarity,
cyan [gray] showing quiet regions with predominantly positive
[negative] polarity, orange indicating sunspots, and green indicating
filaments.}
\label{fig07}
\end{figure*}

Figure \ref{fig07} illustrates the synergistic power of combining
multiple databases.
Stacking up a solar cycle's worth of OMNI wind speeds versus
Carrington longitude reveals the presence of high-speed streams
that recur over multiple rotations and fade in and out over time
\citep[see also][]{Lc09}.
The occurrences of these streams line up quite well with the presence
of large equatorial coronal holes as recorded in the McIntosh
synoptic image archive \citep{Gi17a}.
The long-lived coronal holes (blue/red) seen in panel (b)
are rotating at a rate somewhat faster than the 27.275~day Carrington
rotation, and thus they have a positive slope in this plot.
This correlates well with the slopes seen in the fast wind streams
indicated in panel (a).

The correlations shown in Figure \ref{fig07}
do not stop at the solar wind, but indeed extend
to the Earth's space environment and upper atmosphere.
Clear connections can be found between high-speed solar wind streams
and modulations of the aurora and geomagnetic indices, radiation
belts, ionosphere, and thermosphere \citep{Gi09,So10,Lei11}.
Long-time series analyses over years and decades show periodicities
in all of these quantities that may be associated with periodicities
in the fast solar wind, and consequently the distribution of open
magnetic flux at the Sun in the form of coronal holes \citep{Em11,Lv12}.

The connection between large coronal holes and the fast wind is clear,
but the remaining connections between other coronal structures and the
slow wind are less well understood.
An exact census or mass budget of slow-wind source regions has not
yet been constructed \citep[see also][]{Po13,Ki16,Ab16},
but the following contributors may be significant:

\begin{enumerate}

\item

Steady flows from the {\em boundaries of coronal holes} are often
viewed as the open-field ``legs'' of helmet streamers \citep{WS90,St02}.
When the axis of the streamer belt is oblique to the line of sight,
these structures may be identifiable in coronagraph images merely as
diffuse patches of Quiet Sun.
In either case, the open field lines in these regions tend to expand
more superradially than the central regions of the large coronal holes.
\citet{Sk15} coined the phrase ``boundary wind'' for this component,
which tends to be compositionally similar to the fast wind despite
its lower asymptotic speed.
The smallest coronal holes, which are known to be correlated with slow
wind speeds at 1~AU \citep{No76}, may also be close cousins of these
boundary-layer type flows.

\item

A more time-variable component of the slow wind may be the result of
{\em multi-scale magnetic reconnection} in the corona; i.e., the
opening up of previously closed magnetic loops.
Theoretical arguments for this scenario are discussed below in
Sect.\  \ref{sec:model:debate}.
Evidence for large-scale intermittent mass loss in the HCS (in
the form of low-frequency density fluctuations) was summarized above.
In addition, smaller jet-like reconnection events have been suggested
to feed mass into the solar wind \citep{Mo11,Mj12,Rao16},
especially when they occur near topological boundaries of magnetic
connectivity.
However, \citet{Pa15} concluded that the hot jets seen in X-ray images
convert most of their magnetic energy into heat and not kinetic energy.
Thus, it is unclear whether these reconnection events are powerful or
numerous enough to make a major contribution to the solar wind
\citep[see also][]{Li16}.

\item

Images and spectra of {\em active regions} show rapid flows with
speeds of at least 100 km~s$^{-1}$ along their fanned-out edges
\citep[e.g.,][]{Hr08,BW11,Mo13,ZP16}.
The slow solar wind associated with these structures may come from
small, short-lived coronal holes adjacent to the active regions
themselves \citep{Wa09}.
Active-region slow wind tends to be associated with larger expansion
factors, stronger magnetic fields, higher mass fluxes, higher
O$^{+7}$/O$^{+6}$ ratios, and larger abundance enhancements of low
first ionization potential (FIP) elements than the slow wind
associated with streamers.

\item

Although there is still some debate, it is becoming increasingly
clear that {\em pseudostreamers} are sources of slow solar wind
\citep{RL12,Ck14,Ow14}.
Open field lines near pseudostreamers are topologically complex and
``squashed,'' and the asymptotic speed of their solar wind may depend
on small details of their geometric expansion \citep{Wa12,PV13,Gi17b}.
Nevertheless, the narrow HCS generally appears to be surrounded by
a web-like band of pseudostreamer separatrix surfaces \citep{An11},
and the 20$^{\circ}$ to 30$^{\circ}$ width of this band in latitude
corresponds closely to the zone of slow solar wind seen by
{\em Ulysses} (see Fig.\  \ref{fig02}).
Wind streams associated with pseudostreamers tend to have charge
states and kinetic properties intermediate between those typical of
fast and slow wind \citep{Wa12,Ab15} and extreme values of the
proton mass flux \citep{Zh13}.

\end{enumerate}

Although the coronal magnetic field is not yet measurable in a
routine way, there are several semi-empirical extrapolation models
that have been successful in estimating how the photospheric field
maps out into the heliosphere.
The community's workhorse is the potential-field source-surface (PFSS)
technique, which assumes the corona is current-free between the
photosphere and a spherical surface in the mid-corona, typically at
$r = 2.5 \, R_{\odot}$ \citep{Sc69,AN69}.
Above the so-called source surface, the magnetic field is assumed to be
stretched out by the solar wind into a radially-pointing
``split monopole'' configuration.

\begin{figure*}
\includegraphics[width=1.01\textwidth]{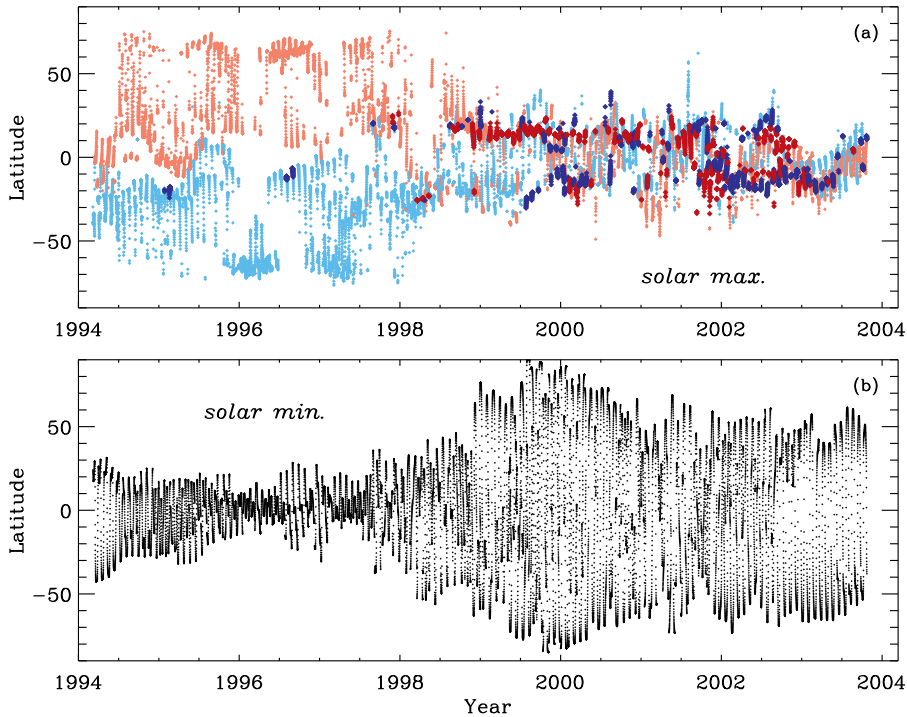}
\caption{(a) Latitudes of photospheric footpoints of open field lines
that connect to the ecliptic plane.
The PFSS technique was used to extrapolate synoptic magnetogram
data from the Wilcox Solar Observatory \citep{HS86}, and a series of
133 sequential Carrington rotations was stacked together in time.
Red [blue] points show footpoints with positive [negative] polarities.
Large points with darker colors indicate strong photospheric fields
($| B_r | > 5$~G), and small points with lighter colors indicate
weak fields below this threshold.
(b) Latitude of the HCS neutral line at the source surface, mapped
from the same set of PFSS models as in panel (a).}
\label{fig08}
\end{figure*}

The PFSS technique is computationally efficient to implement, and it
reproduces a number of large-scale features of the corona as seen with
coronagraphs and during eclipses \citep{Ri06}.
Figure \ref{fig08} shows how PFSS models can also be useful tools
for mapping the origins of solar wind streams
\citep[see also][]{Lu02,Lw04,Fz16}.
At solar minimum, it is clear that high-latitude coronal holes have
significant ``reach'' down into the ecliptic plane.
However, the persistently low latitude of the HCS also means that slow
wind from equatorial streamers must also contribute to the measurement
record at 1~AU.
At solar maximum, the Sun's dominant dipole field is in the process of
being destroyed and reconstituted with opposite polarity, so the tilted
HCS tends to spend time at nearly all latitudes \citep[see also][]{Ri01}.
Interestingly, the distribution of photospheric footpoints of open
field appears to trace out the well-known butterfly diagram of active
regions \citep[see also][and references therein, for discussions of similar
patterns observed in the long-term evolution of coronal holes]{Gi17a}.

\citet{Lv77} and \citet{WS90} found that the asymptotic solar wind
speed along a field line tends to be inversely correlated with the
amount of transverse flux-tube expansion between the photosphere and
a reference point in the mid-corona.
This has been subsequently formalized using the PFSS source surface
at $r = 2.5 \, R_{\odot}$ as the reference point.
The anticorrelation between the wind speed $u$ and the flux-tube
expansion factor $f$ is most evident for the largest structures like
polar coronal holes and low-latitude streamers.
Further refinement in the exact functional dependence of $u$ on $f$ and
other parameters has led to the widely-used Wang-Sheeley-Arge (WSA)
empirical model \citep[see also][]{AP00,Ar03,WS06,Ri15}.

It should be noted that the PFSS technique is only an approximation
to the true three-dimensional structure of the coronal magnetic field.
Stopping short of performing fully global MHD simulations (see below),
there have been a number of attempts to improve on the accuracy of
PFSS-like extrapolation methods.
Many of these methods, along with their alphabet soup of acronyms,
have been reviewed comprehensively by \citet{Su12}.
One noteworthy technique is the so-called current-sheet source-surface
(CSSS) model, which adds some complexity by inserting another spherical
surface between the Sun and the source surface \citep{ZH95}, but also
may provide improvement to solar wind stream prediction \citep{PZ14}.

There has also been substantial effort devoted to improving the
ability of the WSA method to predict the wind speed at 1~AU.
Despite its successes, the time-averaged correlation coefficient
between the predicted and measured wind speed tends to never exceed
$\sim$50\% \citep[e.g.,][]{Mg11a,Gr14,Ri15}.
Statistical comparisons between multiple models and the observed
solar wind \citep{Ji15,Ji16} show generally similar results for other
quantities such as density, temperature, and the magnetic field.
Improvements to the original \citet{WS90} anticorrelation have been
found by including a second parameter, such as the angular distance
$\theta$ between each field-line footpoint and the nearest coronal
hole edge \citep{Ar03,Ow08,Sh12,RL12}
or the magnetic field magnitude at the source surface
\citep{Su06,Fu15,Wa16}.
In addition to magnetic-field parameters, it is possible that data
from EUV images of the chromosphere and low corona can be used
to improve these kinds of empirical predictions
\citep{LM07,Lo08,Rot15}.

Lastly, it is important to note that there is a difference between
the largest-scale stream structure of the solar wind, which
clearly survives the journey to 1~AU, and smaller-scale
structure, which may or may not have a one-to-one correspondence
with features on the Sun.
There have been many reports of {\em in~situ} ``microstreams''
that may be the imprints or relics of coronal structures
\citep{Th90,Rs99,Bv08,Bv16}.
However, there are several stochastic processes that appear to
vigorously blend or scramble the plasma and magnetic flux tubes to
such a degree that deterministic mappings may not be possible.
Sect.\  \ref{sec:model:bimodal} discusses these processes in more
detail.

\section{Physical Processes that Produce the Solar Wind}
\label{sec:model}

Empirically based prediction techniques have been successful,
but it can be argued that moving beyond correlations into the
realm of fundamental physics is required to make substantial new
gains in predictive accuracy.
Once the key physical processes are identified and characterized,
it will be much more straightforward to benchmark, assess, and
refine the simulations used for predicting heliospheric conditions
at 1~AU.
This section reviews recent work along these lines.
Sect.\  \ref{sec:model:agree} begins by outlining the ideas about
which most researchers agree, and Sect.\  \ref{sec:model:debate}
describes the areas of active debate.
Sect.\  \ref{sec:model:bimodal} discusses one notable difficulty in
choosing between the various model proposals: the fact that wind
streams tend to lose their unique connections back to the corona
due to a range of dynamical effects.

\subsection{Uncontroversial Fundamentals}
\label{sec:model:agree}

The Sun's corona is hot.
Although Grotrian, Edl\'{e}n, and others began to understand
the high ionization state of coronal emission lines in the 1930s,
it was left to \citet{A41} to assemble additional lines of evidence
and make the definitive case that the corona is comprised of
plasma with $T \approx 10^6$~K \citep[see also][]{PD14}.
The high gas pressure gradient in such an extended atmosphere led
\citet{P58} to determine that the most likely steady state
would be a supersonic outflow.
This is still the dominant idea in solar wind theory, but there may
be other supplementary sources of radial acceleration in addition
to the gas pressure gradient \citep[see, e.g.,][]{J77,HI02}.
 
The mechanism by which the coronal plasma is heated is not yet known,
but its ultimate energy source is universally understood to be the
convection zone.
Photospheric granulation enables some kind of upward Poynting flux
that delivers kinetic and magnetic energy to the higher layers of
the atmosphere.
After an undetermined time over which much of this energy is ``stored''
in the magnetic field, it is converted irreversibly to heat.
Some of that thermal energy conducts back down to the chromosphere,
and some is extracted by the \citet{P58} mechanism to do work
against the Sun's gravitational potential.
In steady-state, the power input at the base (i.e., energy flux
multiplied by available surface area) should equal the solar wind's
kinetic power far above the solar surface,
\begin{equation}
  \left[ 4\pi r^2 f \left( F_{\rm heat} - F_{\rm cond}
  + \frac{1}{2} \rho u_r^3
  - \rho u_r \frac{GM_{\odot}}{r} \right) \right]_{\rm base}
  \, \approx \,
  \left[ 4\pi r^2 \left( \frac{1}{2} \rho u_{r}^3 \right)
  \right]_{r \gg R_{\odot}}
  \label{eq:fluxes}
\end{equation}
where $f$ is the surface filling factor of magnetic field lines that
eventually reach the solar wind, $F_{\rm heat}$ is the energy flux
deposited by the still-unidentified source of coronal heating,
and $F_{\rm cond}$ is the energy flux conducted back down to the
chromosphere \citep[see also][]{Hm82,HL95,SM03,CS11}.
The equation above neglects enthalpy fluxes and radiative losses,
both of which are usually negligible above the transition region.
From the standpoint of the supersonic solar wind, the high coronal
temperature is only a kind of temporary holding area; i.e., a
stopover between the original source of the energy and its eventual
destiny as outflowing kinetic energy.

The energy balance shown in Equation (\ref{eq:fluxes}) sets the
mass loss rate $\dot{M}$ of the wind, but the relative magnitudes
of the terms on the left-hand side are still not known.
This is an analogous situation to the long-studied problem of
heating in static coronal loops \citep[e.g.,][]{RTV}, in which the
``base pressure'' is determined by time-steady energy conservation.
Both the wind's $\dot{M}$ and a loop's pressure are measures of
how much plasma is drawn up from the relatively vast chromospheric 
reservoir.
A key point is that the mass loss rate is {\em not} determined
by the \citet{P58} solution of the momentum equation.
The accelerating flow through the Parker critical point (i.e.,
the radius at which the wind speed exceeds the sound speed) merely
takes whatever mass is supplied at the coronal base and draws it out.
\citet{Wa98} estimated the sphere-averaged value of $\dot{M}$ varies
between about $2 \times 10^{-14} \,\, M_{\odot} \mbox{yr}^{-1}$
(at solar minimum) to $3 \times 10^{-14} \,\, M_{\odot} \mbox{yr}^{-1}$
(at solar maximum).

In the past, theorists have disagreed about whether the solar wind
is more properly described using fluid or kinetic equations
\citep{Ch60,Jo70,LS71}.
The consensus now is that both pictures agree on the basic properties
of the outflow \citep{LP01,P10}.
To some extent, this ought to be the case, because the conservation
equations based on fluid moments are derived directly from
Liouville's theorem and the associated kinetic transport equations.
However, the fluid picture does make closure assumptions about
the shapes of the velocity distribution functions, and there remain
disagreements about, e.g., the validity of classical heat conduction
\citep{LP03} and the available linear wave modes \citep{Vs17}.

It has also been proposed that there are suprathermal particles
(i.e., power-law tails that augment the normally Maxwellian
velocity distributions) in the solar atmosphere, and that these
particles escape preferentially to produce high coronal temperatures
\citep{Lv74,Sc92}.
This ``velocity filtration'' idea has been implemented in kinetic
exobase-type models that successfully predict some aspects of the
particle measurements at 1~AU \citep[e.g.,][]{MV99,Zo04,PP14}.
Despite this idea being somewhat outside the mainstream of research,
we list it here in the subsection about uncontroversial physics.
It may or may not be important on the Sun, but it is similar to any
other coronal heating theory in that it requires converting some other
form of energy (i.e., kinetic or magnetic) into thermal energy.
The difference is that this conversion would have to occur down in the
chromosphere, where a combination of Coulomb collisions and radiative
losses would keep the majority of particles cool.

A final uncontroversial statement to make about the solar wind is
that it its fluctuations (e.g., waves, turbulence, shocks, and
end-products of magnetic reconnection) are likely to both
affect and be affected by the time-averaged properties of the flow.
There do not seem to be any theories of coronal heating and solar
wind acceleration that do {\em not} ultimately involve the summed
impact from multiple transient or oscillating events.
The extent to which terms like ``waves'' and ``turbulence'' are
useful descriptors of the physics is still being debated, but
the variability is ubiquitous and important.

\subsection{Controversial Alternatives}
\label{sec:model:debate}

The exact chain of events by which the corona is heated and the
solar wind is accelerated is not yet known.
It has proven exceedingly difficult to distinguish between competing
theoretical models because the basic energy conversion processes appear
to be acting on spatial and time scales unresolved by existing
observations.
It is also probably the case that different mechanisms are dominant
in different source-regions of the solar wind (e.g., active regions
versus coronal holes), and that in some regions multiple mechanisms
may be contributing at comparable levels.

With the above caveats in mind, we have sorted the proposed physical
models into three broad categories:

\begin{enumerate}

\item
If solar wind field lines are open to interplanetary space---and if
they remain open on timescales comparable to the time it takes plasma
to accelerate into the corona---then the main sources of energy must
be injected at the footpoints.
Thus, in {\em wave/turbulence-driven (WTD) models,}
the convection-driven jostling of the flux-tube is assumed to
generate wave-like fluctuations that propagate up into
the extended corona
\citep{C68,H86,Ve91,Mt99,SI06,CvB07,Of10,vB11,Ch11,MS12,PC13,Li14,TV17,vB17}.
The coronal heating comes from wave dissipation, whose physical origin
is still a subject of debate.
Fast and slow wind streams come from the fact that flux tubes with
different expansion factors have different radial distributions of
the heating rate and different locations of the Parker critical point
\citep[see, e.g.,][]{LH80,Cr05}.

\item
Near the Sun, all open magnetic flux tubes are observed to exist in
the vicinity of closed loops.
The complex distribution of mixed-polarity loop footpoints---which
is evolving continuously via emergence, cancellation, diffusion,
splitting, and merging---has been called the Sun's ``magnetic carpet''
\citep{TS98}.
It is natural to propose a class of
{\em reconnection/loop-opening (RLO) models,} in which the mass and
energy in some coronal loops is fed into the open regions
that connect to the solar wind.
Some have suggested that RLO-type energy interchange primarily occurs
at the scale of the supergranular network
\citep{Ax92,Fi99,Fi03,Sw06,Ya13,Kp17}, and others favor
larger-scale reconnection events near global null points and streamer
cusps \citep{Su96,Ei99,Wa00}.
The idea of a so-called S-web, or separatrix-web \citep{An11,Em12,Hg17}
involves a continuous range of scales between the two, with the
complex topological rearrangement helping to energize the slow wind.

\item
The upper chromosphere is filled with a range of narrow features
known variously as spicules, jets, fibrils, surges, and mottles.
Some have suggested that much of the corona's mass and energy may
be injected directly from these structures
\citep{Pn86,Lo94,DP09,Mo11,Ti14,Rao16}.
Strictly speaking, this idea could be considered a subset of either
the WTD or RLO models, depending on whether the spicules and jets are
driven by waves \citep{SH84,KS99,CW15} or by reconnection
\citep[e.g.,][]{Uc69,Pr16}.
Still, the direct chromospheric source of the mass appears to
distinguish this idea---here called {\em chromospheric mass supply
(CMS)}---from the other two, in which the processes giving rise to
the corona and solar wind are generally located up in the corona itself.

\end{enumerate}

The above list of processes does not include some that have been
applied mostly to closed loops that do not connect directly to the
solar wind.
For example, the classical idea of direct-current (DC) heating---in
which the corona evolves in a succession of twisted and quasi-static
states via small-scale current dissipation events
\citep{P72,HP84,vB86}---does not sit solidly within the WTD,
RLO, or CMS categories.
These DC mechanisms are often associated with {\em nanoflares:}
tiny episodic bursts of energy that may dominate the coronal heating
\citep{P88,PJ00,Jo16}.
It is fair to say that all three of the above model
categories can produce nanoflare-like intermittent heating.
In many cases, however, the predicted small-scale bursts are expected
to be highly dynamic and not quasi-static; this is consistent with
observations of ``braiding'' by high-resolution imagers
\citep[e.g.,][]{Ci13,vB14}.
No matter the source of these bursts, they may generate nonthermal
electrons that propagate out into the heliosphere \citep[e.g.,][]{CG14}.

Observations have been used to attempt to support or rule out the
above processes, but no consensus has been reached.
For example, \citet{Rb10} claimed there to be insufficient WTD
energy present to heat the corona.
\citet{CvB10} similarly claimed the RLO model cannot energize the
plasma in open-field regions \citep[see also][]{KP11,Li16}.
\citet{KB14} concluded that CMS-type processes cannot produce sufficient
heat to power the corona.
However, none of these ideas has been ruled out conclusively because
(1) the relevant energy-release events cannot yet be measured directly,
and (2) most models still employ free parameters that can be adjusted
to improve agreement with the existing data.

As mentioned above, it may also be possible for aspects of more than
one model to be present.
\citet{Sk16} suggested that both closed loops and open
flux tubes produce solar wind---with closed loops contributing more to
the slow wind and open flux tubes contributing more to the fast
wind---and that once the plasma is released, both types share a single
WTD acceleration mechanism.
There are other ways that the WTD, RLO, or CMS models can exhibit
intermingled characteristics.
If there is a turbulent cascade that produces WTD-type heating,
the ultimate energy dissipation may be best describable by small-scale
reconnection events that reconfigure the magnetic topology
\citep[e.g.,][]{MV11}.
On the other hand, the proposed RLO reconnection events in the magnetic
carpet may generate MHD waves \citep{Ly14,Kp17} that go on to dissipate
and heat the plasma.
Lastly, even if the evolving S-web does not produce sufficient
reconnection to heat the global corona, it may be that the mere
presence of the sharp transverse gradients and separatrices can act
as an additional source of shear-driven waves
\citep[see, e.g.,][]{LR86,Kg99,Kg07,Ev12}.

How can we as a community progress toward the goal of identifying
and characterizing the physical processes at work in the solar corona?
It is worthwhile listing some promising paths forward (see
Sect.\  \ref{sec:paths}), but
there have also been unsuccessful attempts to rule out models.
For example, the prevalence of enhanced low-FIP elements and high
freezing-in temperatures (e.g., high O$^{+7}$/O$^{+6}$ ratios) in the
slow wind has been used as evidence for RLO-type processes.
Closed coronal loops exhibit similar composition patterns to the slow
wind, so it is natural to connect them together.
However, time-steady WTD models have been shown to predict variations
with wind speed in abundances and charge states that follow the measured
patterns in interplanetary space \citep{CvB07,Jin12,Cr14a,Or15}.
Thus, for now, it appears that solar wind composition is not a
useful discriminator between the main theoretical categories.

\subsection{Does Radial Evolution Mask Bimodality?}
\label{sec:model:bimodal}

It is still not clear whether the dominant acceleration processes
in the corona produce {\em bimodal} (i.e., cleanly separated) fast
and slow wind streams, or if they produce a continuous distribution
of states by varying one of more parameters.
In some versions of the WTD model, a slow variation of the superradial
flux-tube expansion factor can produce a bimodal jump in the wind
speed \citep{Cr05}.
This occurs because there are often multiple possible radii for the
\citet{P58} critical point, and the global time-steady solution
can undergo a rapid transition from one of those radii to another,
depending on small changes in the expansion factor
\citep[see also][]{KH76}.
If the overall radial distribution of coronal heating remains
more or less unchanged, a wind with a lower critical point tends
to have a higher asymptotic speed, and a wind with a higher critical
point tends to have a lower speed \citep{LH80,Pn80}.

Whether or not the corona produces a bimodal or broad/continuous
distribution of wind speeds, it remains difficult to use {\em in~situ}
interplanetary data to make definitive conclusions about this issue.
In the ecliptic plane, the solar wind becomes mixed---in ways
that usually increase its randomness and stochasticity---in
at least three distinct ways.
(1) MHD turbulence gives rise to an effective random walk of
field lines and a potential loss of identity for initial plasma
parcels that become shredded in space and time \citep{Mt98,Gr12}.
(2) The Sun's rotation produces stream-stream interactions that
create CIR spiral structures and ever-greater longitudinal
blending with increasing distance \citep{BS99,Ri07}.
(3) Coulomb collisions, whose randomizing effects accumulate with
radial distance, tend to erase the field-aligned kinetic effects 
discussed in Sect.\  \ref{sec:obs} \citep[e.g.,][]{Ka08}.
The main impact of these processes is to produce ambiguity when
mapping wind streams back from interplanetary space to the Sun.
This ambiguity becomes more apparent for smaller scales in longitude
and latitude, and for more distant mappings from the outer heliosphere
\citep[e.g.,][]{El12}.

\begin{figure*}
\hspace*{0.015in}
\includegraphics[width=1.008\textwidth]{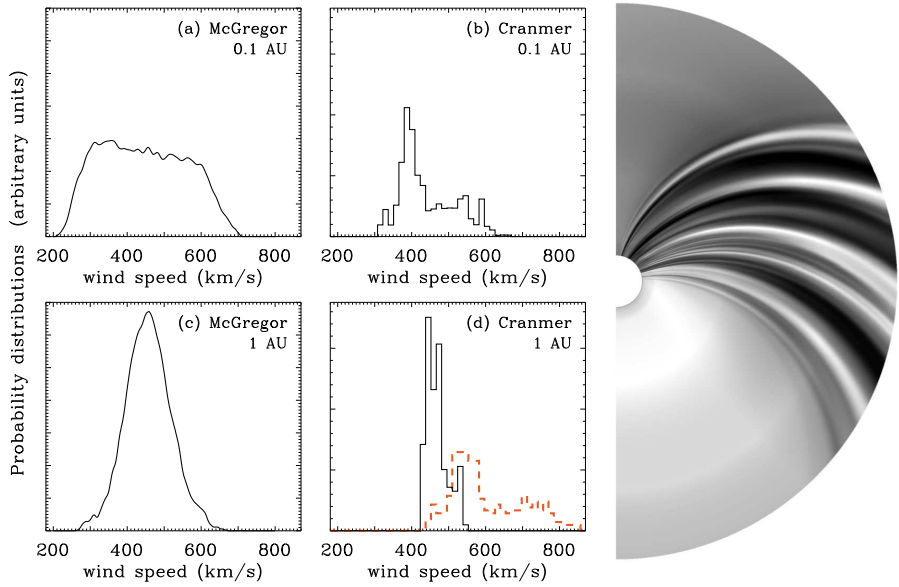}
\caption{Comparison of solar wind-speed probability distributions in
near-ecliptic regions of semi-empirical three-dimensional models:
(a,c) WSA-ENLIL models \citep{Mg11b} assembled over 3 years during the
1995--1997 solar minimum, sampled at 0.1~AU and 1~AU;
(b,d) coordinated set of ZEPHYR models \citep{CvB13} for a quiet
region observed with SOLIS in 2003, sampled at 0.1~AU and 1~AU.
The red dashed curve is the distribution of wind speeds at 1~AU
for separate ZEPHYR-model flux tubes computed {\em without} stream
interactions in the ecliptic plane.
Right-hand panel shows the simulated CIR pattern for the 2003
quiet region. Grayscale levels correspond to density variations,
with the large-scale radial dropoff removed.}
\label{fig09}
\end{figure*}

Figure \ref{fig09} illustrates the second of the three mixing
processes listed above.
Model-based reconstructions of radial and longitudinal evolution
of the solar wind are shown from \citet{Mg11b} and \citet{CvB13}.
High-contrast flux-tube structure near the Sun appears to be
eroded rapidly by CIR stream interactions.
The most extreme solar wind parcels (e.g., the highest and lowest
speeds) tend to disappear and leave behind a single-peaked distribution
dominated by moderate speeds of order 400--450 km~s$^{-1}$.
The red dashed curve in Figure \ref{fig09}d shows what the
distribution of speeds would have been if stream-stream interactions
in the ecliptic were ignored (i.e., the slight bimodality at 0.1~AU
would have been preserved at 1~AU).
Thus, the presence of stream-stream interactions tends make any
intrinsic coronal bimodality extremely difficult to detect at 1~AU.

\section{Paths Forward to Improved Space Weather Prediction}
\label{sec:paths}

Although past observational and theoretical work has improved
our understanding of the basic processes at work in generating the
solar wind, much more needs to be done.
The subsections below describe some of what is on the horizon for
pure theory (Sect.\  \ref{sec:paths:theory}),
observations and {\em in~situ} measurements
(Sect.\  \ref{sec:paths:obs}),
and empirical prediction tools that attempt to make use of the best
of what models and data have to offer
(Sect.\  \ref{sec:paths:emp}).

\subsection{Theoretical Improvements}
\label{sec:paths:theory}

In order to determine the quantitative contributions of WTD, RLO,
and CMS processes to accelerating the solar wind, each of these models
must be developed to the point of eliminating their free parameters.
The difficulty in doing this lies in the large range of relevant
spatial and time scales important to the physics---e.g., the Sun can
exhibit wavelike variability with periods from milliseconds \citep{Ba98}
to years \citep{Mc17}.
This reality demands adaptive, multi-scale modeling techniques that
go beyond many traditional plasma-in-a-box type simulations.
These models must also strive to contain as many of the proposed
physical processes as possible.
The true impact of any one process on the system may not be made clear
until it is allowed to interact with the others in a realistic way.

Although there are still many physics questions that can be answered by
models with limited spatial extent, the community's ultimate goal is to
develop and improve {\em global three-dimensional simulations}
of the entire corona and heliosphere.
In the last decade, several of these models have made the transition
from polytropic energy equations and prescribed
heating functions to more physics-based (usually WTD) ways of
computing the coronal heating.
Additional recent developments include solving multi-fluid energy
equations instead of single-fluid MHD \citep{Us12,Us16,vdH14,Or15}
and assimilating time-dependent photospheric data instead of using
static synoptic maps \citep{Fe15,Ln16,Ya17}.
Aspects of the simulations that have been shown to be especially
important for space weather prediction include
resolving the sharp HCS \citep[i.e., avoiding unphysical diffusion
associated with too coarse a grid;][]{St12} and using Monte Carlo
ensembles instead of single models \citep{Ri13,Ow17}.

Because of the need to compare model predictions with actual
observational data (see below), it is important to include
{\em forward modeling} in the theorist's toolbox.
It has long been a goal to ``invert'' the data; i.e., to extract
information from images and spectra that allow us to solve for the
three-dimensional distributions of plasma parameters such as
temperature and density.
However, the corona is optically thin, highly structured, and
time-variable.
Thus, attempts at data inversion are often nonunique and fraught with
uncertainty \citep[see, e.g.,][]{JM99,DB08}.
It is a much safer procedure to take the theoretical model output,
simulate what observers would see, and make direct comparisons at
the level of the data.
Several sets of sophisticated software tools are being developed
to enable this kind of forward modeling
\citep[e.g.,][]{Ni15,Gi15,Gi16,VD16}.
Methodologies are also being developed to iteratively optimize models
to match observations, thus achieving the ultimate goal of inversion
\citep[e.g.,][]{Da16}.
A benefit of these approaches is that they provide information about
which observables are the most influential in validating or falsifying
a given model.

\subsection{Observational Improvements}
\label{sec:paths:obs}

Despite the difficulties described above regarding forward and
inverse modeling, we need to remain on the lookout for new ways in
which observations can put tight constraints on theory.
A clear source of ``low-hanging fruit'' is to pursue better
measurements of the ingredients of proposed physical mechanisms.
For example, testing the WTD model requires knowing the amplitudes and
damping rates of fluctuations that propagate along open field lines
(i.e., filling in the gaps in Figure \ref{fig06}).
Testing the CMS model requires knowing how much mass and Poynting
flux comes up through the photosphere, chromosphere, and transition
region---and does not come back down again.
Testing the RLO model requires measuring how much plasma and magnetic
energy gets processed through magnetic reconnection events that
convert closed to open field lines.

Global models of the solar wind depend on
photospheric magnetic field maps as lower boundary conditions.
Traditional synoptic maps, built up from longitudinal magnetograms,
are problematic because (1) they do not contain information about
magnetic currents in the solar atmosphere, (2) they ignore variability
on timescales shorter than a solar rotation, and (3) the north and south
poles are poorly resolved \citep{Su11,Pt17}.
Vector magnetograms are improving the situation \citep[e.g.,][]{Liu17},
but it is still the case that magnetograms from different observatories
tend to produce markedly different predictions for the
ecliptic plane at 1~AU \citep{Ji11,Ri14}.
The ideal solution would be to have telescopes on multiple spacecraft,
positioned throughout the heliosphere, so that all $4\pi$ steradians
of the Sun can be monitored continuously \citep{Ro04,Lw08,Sr12}.
A continuous view of the solar poles is particularly compelling, both
for space-weather monitoring and for establishing the nature of the
solar dynamo.
There are also mission concepts that would provide new insights
while stopping short of full $4\pi$ coverage, such as an
early-warning system at the Earth--Sun L5 point
\citep[e.g.,][]{Lv16,Pv16}, or a two-spacecraft system that would
improve upon {\em STEREO's} initial exploration of stereoscopic
imaging \citep{Sg15}.
Sustained multi-vantage observations of the photospheric magnetic field,
the inner boundary of the heliosphere, has transformative potential.

Having the ability to make direct and routine measurements of the
coronal magnetic field would complement the photospheric data
\citep{Lin04,Ju13}.
Most proposed coronal heating mechanisms are magnetic in nature, and
they often depend on the properties of twisted, non-potential structures
that are difficult to extrapolate up from photospheric boundary conditions.
Indirect methods such as coronal seismology \citep[e.g.,][]{DN12}
have been helpful, but direct measurements of the field magnitude
and direction could put more stringent constraints on models.
Observations of linear polarization in infrared coronal emission lines
by the {\em Coronal Multichannel Spectropolarimeter} \citep{Tz08}
have demonstrated the power of such observations for establishing the
topologies of magnetic structures such as flux ropes and pseudostreamers
\citep{Dv11,BS13,Rm13,Rm14,Gi17b}.
The {\em Daniel K.\  Inouye Solar Telescope}
\citep[DKIST;][]{Tri16} will provide a more direct measure of the
line-of-sight coronal field strength via high-precision measurements of
circular polarization, and explore new regimes of coronal magnetometry.
Future programs, such as the
{\em Coronal Solar Magnetism Observatory} \citep[COSMO;][]{Tz16},
would enable full-Sun, synoptic measurements of the coronal magnetic field.
Ultimately, a vantage away from the Earth-Sun line would be beneficial
for monitoring Earth-directed space weather; the proposed
balloon-borne {\em Waves and Magnetism in the Solar
Atmosphere} \citep[WAMIS;][]{Ko16} and the
{\em massively-multiplexed Coronal Spectropolarimetric Magnetometer}
\citep[mxCSM;][]{Lin16} represent development in this direction.

Additional information about solar wind acceleration is likely to be
learned by detecting scattered light from plasma parcels in the extended
corona ($r \approx 2$--30 $R_{\odot}$).
This range of heights is where coronal ``structures'' evolve into
solar wind streams.
It is also where the plasma becomes collisionless; i.e., where
departures from thermal equilibrium (Fig.\  \ref{fig05}) start to
become useful probes of the physics.
It would be beneficial for next-generation ultraviolet coronagraph
spectrometers \citep[e.g.,][]{Ko08,St16} to be developed, in order
to follow up on the successes of UVCS and extend the remote-sensing
field of view to larger heights and more ions.

To monitor solar wind acceleration at slightly larger distances,
there are opportunities for improving both the analysis techniques
and instrumentation for space-based visible-light
heliospheric imagers \citep{Ru11,Df11,Df16,DH15} and
ground-based radio arrays \citep{Mn17}.
Large-scale collaborations such as HELCATS \citep[Heliospheric
Cataloguing, Analysis, and Techniques Service;][]{Pk16,Ru17}
are building a broad range of analysis tools and observational
databases.
New remote-sensing heliospheric observations have the potential to
improve our knowledge of how parcels accelerate in both isolated wind
streams and CIRs (as well as CMEs), and also to reveal how coronal
waves evolve into turbulent eddies.

Of course, future improvements in solar wind measurements must also
include {\em in~situ} particle and field detection.
We anticipate that the {\em Parker Solar Probe} \citep{Fx16}
and {\em Solar Orbiter} \citep{Mu13}, both scheduled for launch in 2018,
will revolutionize our conception of the inner heliosphere.
Specifically, these missions will explore regions in which all three
mixing processes discussed in Sect.\  \ref{sec:model:bimodal} have not
yet had time to smear out the unique field-line mappings back to the
corona.
As these missions are starting to explore the inner heliosphere,
India's {\em Aditya-L1} mission is expected to launch around
2020.  Its combination of remote-sensing and {\em in~situ}
instruments \citep{Gh16,Vn17} should extend the multi-diagnostic
capabilities pioneered by {\em SOHO} and {\em ACE} at Earth--Sun L1.
Lastly, the {\em Turbulence Heating Observer} \citep[THOR;][]{Vv16}
spacecraft is expected to launch in 2026, and it will be dedicated
to a complete characterization of micro-scale turbulent dissipation in
the solar wind and Earth's magnetosphere.

\subsection{Improvements in Empirical Prediction Tools}
\label{sec:paths:emp}

As discussed in Sect.\  \ref{sec:obs:mapping}, empirically based
techniques to predict solar wind conditions at 1~AU (e.g., PFSS
and WSA) continue to be tested and upgraded.
Comparisons are being made between potential-based field
extrapolations and global MHD models \citep{Ji15,Ji16}, but both
techniques still have problems reproducing the full range of
observed solar-wind variability.
Ongoing improvements in model validation are being made by
collaborative efforts such as
the Community Coordinated Modeling Center \citep[CCMC;][]{Hs10},
the Space Weather Modeling Framework \citep[SWMF;][]{To05},
the European Heliospheric Forecasting Information Asset
\citep[EUHFORIA;][]{Pm17}, and
the NOAA Space Weather Prediction Center \citep[SWPC;][]{Bg15}.

Eventually, forecasts based on solutions to conservation equations
ought to replace empirical recipes such as PFSS, but at present
that appears to be too computationally expensive.
A promising middle ground may be to use extrapolations based on
{\em magnetofrictional evolution,} which stops short of full MHD but
still manages to include the time-dependent development of non-potential
coronal currents \citep{Ye14,Ed15,Fi15}.
For the actual prediction of solar wind properties along open field
lines, there are new semi-empirical tools \citep[see, e.g.,][]{WC14,PR17}
that use the output of physics-based models (rather than WSA-type
correlations) but are also designed for computational speed.
For solar-cycle-length predictions, it may be important to couple
corona/wind models to flux-transport models of the photospheric fields
\citep[e.g.,][]{Mr16,Wz16} or even to more self-consistent simulations
of the interior dynamo.

Space weather forecasting also requires efficient {\em data assimilation}
\citep[or {\em data incorporation,} as discussed by][]{Sj15}
for it to be accurate.
Unfortunately, there is a stark contrast with terrestrial weather
forecasting, in that (1) the space-based data are so sparse that the
``upwind conditions'' are not really known with sufficient precision, and
(2) the corona/heliosphere physics that produces these upwind conditions
is still not understood.
There are also differences in how data products are used.
Photospheric magnetograms are straightforward drivers of the models
because they can be used as lower boundary conditions.
However, it is more difficult to use data measured in the corona and
inner heliosphere, because those regions are inside the computational
grids of most MHD simulations.
Just straightforwardly inserting newly measured plasma properties into
an MHD model would overconstrain the equations and produce discontinuities
in the conserved quantities.
One promising path forward is to run multiple simulations in
parallel---with randomly varied initial/boundary conditions or different
choices for unknown coronal heating processes---and then to choose the
model (or linear combination of models) that best matches the data
\citep[see][]{Dx07}.
The optimization method of \citet{Da16} discussed above also suggests
possible solutions to this problem.
This type of {\em ensemble modeling} is now being applied to a range of
solar wind and ICME prediction methods \citep{Lc13,Ri13,Cs15,Ow17}.

\section{Conclusions}
\label{sec:conc}

The objective of this paper has been to review our present-day
understanding of the origins of ambient solar wind streams.
We also examined a wide range of possible ways to improve
space-weather forecasts related to the ambient wind.
There are still some fundamental questions about solar wind
acceleration for which we do not yet have answers:
What are the physical processes that dominate coronal heating?
Where does most of the slow wind come from on the Sun?
Do fast and slow wind streams originate from two separate
mechanisms?
However, it is nevertheless the case that forecasts are improving.
This is due in part to the successes of empirical correlation
techniques, which help point the way to identifying the ``best''
physical processes to include in simulations.
As these processes are identified and characterized, we also expect
improvements in models of CMEs, which have proposed heating
mechanisms \citep[see, e.g.,][]{Liu06,Mu11} very similar to those
discussed in Sect.\  \ref{sec:model} for the ambient wind.

Much of the discussion in this paper has been focused on the
idea of applying insights from fundamental research to the practical
goals of space weather forecasting.
This phase of work has been called ``research-to-operations'' (R2O).
However, there are also benefits that flow in the opposite
direction; i.e., ``operations-to-research'' \citep[O2R;][]{Sb14}.
Increased knowledge about our well-studied Sun and heliosphere 
feeds back into astrophysical studies of Sun-like stars
\citep[e.g.,][]{HS96,Sm03,Br15}
and even more distant objects such as supermassive
black holes \citep{QG99,Li17} and galaxy clusters \citep{Pr12}.
For decades, the Sun has been considered a testbed for studying
fundamental physical processes that cannot be reproduced in
terrestrial laboratories.
Because the solar wind unifies studies of waves, turbulence, and
magnetic reconnection, it also feeds into interdisciplinary studies
of universal processes in heliophysics \citep[see][]{Da09,Ra10}.

\begin{acknowledgements}

The authors would like to thank Ruedi von Steiger, Andr\'{e} Balogh,
Dan Baker, Tam\'{a}s Gombosi, Hannu Koskinen, and Astrid Veronig
for convening the fantastic 2016 ISSI workshop on the scientific
foundations of space weather.
SRC's work was supported by NASA grants {NNX\-15\-AW33G} and
{NNX\-16\-AG87G}, NSF grants 1540094 (SHINE) and 1613207 (AAG), and
start-up funds from the Department of Astrophysical and Planetary
Sciences at the University of Colorado Boulder.
PR's work was supported through a grant from NASA's Living With a
Star (LWS) Program.
The National Center for Atmospheric Research (NCAR) is supported by
the National Science Foundation.
This research made extensive use of NASA's Astrophysics Data System (ADS).

\end{acknowledgements}

\end{document}